\documentclass[utf8]{FrontiersinHarvard}
\usepackage{url,hyperref,lineno,microtype,subcaption}
\usepackage{arydshln}
\usepackage[onehalfspacing]{setspace}
\usepackage[T1]{fontenc}
    
\def\keyFont{\fontsize{8}{11}\helveticabold}
\def\firstAuthorLast{Ganguly {et~al.}} 
\def\Authors{Shalini Ganguly\,$^{1}$, Yuan Li\,$^{1,*}$, Valeria Olivares\,$^{2}$, Yuanyuan Su\,$^{2}$, Francoise Combes\,$^{3}$, Sampadaa Prakash\,$^{1}$, Stephen Hamer\,$^{4}$, Pierre Guillard\,$^{5,6}$, and Trung Ha\,$^{1}$}
\def\Address{$^{1}$Department of Physics, University of North Texas, Denton, TX, USA. \\
	$^{2}$Department of Physics and Astronomy, University of Kentucky, Lexington, KY, USA.\\
 $^{3}$ LERMA, Observatoire de Paris, PSL Research Univ., CNRS, Sorbonne Univ., 75014 Paris, France.\\
 $^{4}$Department of Physics, University of Bath, Claverton Down, BA2 7AY, UK.\\
 $^{5}$Sorbonne Universit\'{e}, CNRS, UMR7095, Institut d’Astrophysique de Paris, 98 bis Bd Arago, 75014 Paris, France.\\
 $^{6}$Institut Universitaire de France, Minist\'{e}re de l'Education Nationale, de l'Enseignement Sup\'{e}rieur et de la Recherche, 1 rue Descartes, 75231 Paris Cedex 05, France.}
\def\corrAuthor{Yuan Li}
\def\corrEmail{yuan.li@unt.edu}

\usepackage{amsfonts,amsmath,amssymb}
\usepackage{graphicx}
\graphicspath{{./}{images/}}

\newcommand{\apj}{ApJ}
\newcommand{\apjl}{ApJL}
\newcommand{\apjs}{ApJS}
\newcommand{\mnras}{MNRAS}

\newcommand{\araa}{ARAA}

\newcommand{\nat}{Nature}

\newcommand{\aap}{A\&A}

\newcommand{\pasj}{Publ. Astron. Soc. Japan}

\begin{document}
    \onecolumn
	\firstpage{1}

	\title[Cluster Filaments]{The Nature of the Motions of Multiphase Filaments in the Centers of Galaxy Clusters} 
	\author[\firstAuthorLast ]{\Authors} 
	\address{} 
	\correspondance{} 
	
	\extraAuth{}
\maketitle
\begin{abstract}
\section{}
The intracluster medium (ICM) in the centers of galaxy clusters is heavily influenced by the ``feedback'' from supermassive black holes (SMBHs). Feedback can drive turbulence in the ICM and turbulent dissipation can potentially be an important source of heating. Due to the limited spatial and spectral resolutions of X-ray telescopes, direct observations of turbulence in the hot ICM have been challenging. Recently, we developed a new method to measure turbulence in the ICM using multiphase filaments as tracers. These filaments are ubiquitous in cluster centers and can be observed at very high resolution using optical and radio telescopes. We study the kinematics of the filaments by measuring their velocity structure functions (VSFs) over a wide range of scales in the centers of $\sim 10$ galaxy clusters. We find features of the VSFs that correlate with the SMBHs activities, suggesting that SMBHs are the main driver of gas motions in the centers of galaxy clusters. In all systems, the VSF is steeper than the classical Kolmogorov expectation and the slopes vary from system to system. One theoretical explanation is that the VSFs we have measured so far mostly reflect the motion of the driver (jets and bubbles) rather than the cascade of turbulence. We show that in Abell 1795, the VSF of the outer filaments far from the SMBH flattens on small scales to a Kolmogorov slope, suggesting that the cascade is only detectable farther out with the current telescope resolution. The level of turbulent heating computed at small scales is typically an order of magnitude lower than that estimated at the driving scale. Even though SMBH feedback heavily influences the kinematics of the ICM in cluster centers, the level of turbulence it drives is rather low, and turbulent heating can only offset $\lesssim10\%$ of the cooling loss, consistent with the findings of numerical simulations.
\tiny
\keyFont{ \section{Keywords:} Galaxy Clusters, Turbulence, Intracluster Medium, Galaxy Physics, Active Galactic Nuclei, X-ray cavities}
\end{abstract}
	
\section{Introduction} \label{sec:intro}

Most relaxed clusters harbor cool cores. In the absence of an effective heating source, the high radiative cooling rate of the gas is expected to lead to a cooling flow of 100s~$\rm M_\odot$~yr~$^{-1}$ onto the Brightest Cluster Galaxies (BCGs) \citep{Fabian1994}. Both observations and theoretical models suggest that such a classical cooling flow is suppressed by the feedback from the active galactic nuclei (AGNs) powered by supermassive black holes (SMBHs) at the centers of these systems. The jets and/or outflows from the SMBHs lead to the formation of shocks and bubbles (cavities) in the intracluster medium (ICM) that can be detected in X-ray and radio. These feedback processes can inject thermal and mechanical energies into the ICM, leading to a quasi-static hot atmosphere where cooling and heating are balanced on average \citep[and references therein]{McNamara2007, Fabian2012}. Exactly how the energy from SMBHs becomes coupled to the ICM is not well-understood. The dissipation of jet-driven turbulence is one of the proposed heating mechanisms that have been discussed extensively in recent years. 
 
Turbulent motion in the hot ICM has been probed using X-ray observations \citep[e.g.,][]{2011MNRAS.410.1797S, 2020A&A...633A..42S, 2022MNRAS.511.4511G, 2022MNRAS.513.1932G, 2023MNRAS.518.2954D}. For example, surface brightness fluctuations observed by \textit{Chandra} can be used to measure turbulence assuming that velocity fluctuations are proportional to density fluctuations \citep{Zhuravleva2014}. In addition, Hitomi has measured X-ray line widths in the central regions of the Perseus Cluster \citep{Hitomi2016}. Although these X-ray studies have limited spatial and spectral resolutions, they consistently suggest that cluster cores are turbulent, and that the inferred level of heating from turbulent dissipation can balance radiative cooling in many systems \citep{Zhuravleva2014, Zhuravleva2018}. 

On the other hand, numerical models of AGN feedback have generally found a low level of turbulence or turbulent dissipation. These include both plane-parallel models \citep{Reynolds2015, Bambic2018} and more realistic setups \citep{Weinberg2017, Prasad2018}. Heating of the ICM is usually achieved via shock dissipation \citep{Li2017}, adiabatic processes \citep{Yang2016}, sound wave dissipation \citep{Bambic2019}, and/or mixing \citep{Hillel2016}. Interestingly, these simulations often produce a velocity dispersion consistent with the Hitomi results on similar spatial scales (tens of kpc scales) \citep[e.g.,][]{Gaspari2018, Prasad2018}. 
    
Multiwavelength observations show that cool-core cluster centers are often multiphase. Cool ionized gas can be observed in the H$\alpha$ using optical telescopes \citep{2010ApJ...721.1262M, Hamer2016}. Several theories have been proposed whereby the hot ICM becomes thermally unstable and condenses to form cool clouds aided by jet uplifting and turbulence \citep{Li2014, McNamara2016, Voit2018}. The cool gas fuels AGN activities, which in turn provides feedback to the ICM. Observationally, cool filaments are often spatially associated with soft X-ray features and some of the cool filaments are found to be at the edge of the X-ray bubbles \citep{Werner2014, Fabian2016}. Some of the cool H$\alpha$ filaments also have a molecular component. CO observations reveal large reservoirs of cold molecular gas with masses of $\sim10^9-10^{11}$~$\rm M_\odot$ in many cool-core clusters \citep{2001MNRAS.328..762E, Salome2003, 2019MNRAS.490.3025R}.

Cluster center multiphase filaments have been observed in great detail in recent years with optical telescopes equipped with Integral Field Units (IFUs), such as the Multi-Unit Spectroscopic Explorer (MUSE), as well as radio telescopes such as the Atacama Large Millimeter/submillimeter Array (ALMA). These high-resolution observations reveal complex kinematics of the filaments on small scales, and generally a lack of ordered motion on large scales, suggesting that their motions may be turbulent \citep[e.g.,][]{Tremblay2018}. The velocity dispersion of the cool filaments on large scales is generally consistent with that of the hot ICM inferred from X-ray observations, suggesting that the kinematics of different phases of the ICM are coupled on large scales \citep{Gendron-Marsolais2018}.
    
Recently, \cite{Li2020} have proposed a new method to study the kinematics of multiphase filaments in galaxy clusters. They compute the first-order velocity structure function (VSF) of the filaments in three nearby galaxy clusters: Perseus, Abell 2597 and Virgo, and find their motion to be turbulent. The inferred driving scales are found to be roughly the sizes of the X-ray bubbles, suggesting that the turbulent motion is driven by SMBH feedback. Within the limited spatial scale that can be probed with \textit{Chandra} X-ray observations, turbulence traced with cool filaments is consistent with the X-ray measurements \citep{Zhuravleva2014, Hitomi2016}. However, the VSFs of all three systems show steeper slopes than the classical Kolmogorov expectation. The exact reason for this is unclear. Proposed theoretical explanations include magnetic fields, shocks, gravity waves, and plasma instabilities \citep{Wang2021, Mohapatra2022, Hu2022a, Arzamasskiy2022}. It may also simply imply that we are mostly tracing the bulk motion of the jets and the jet-inflated bubbles, rather than the cascade of turbulence \citep{Zhang2022}. If we assume that cool filaments are still good kinematic tracers of the hot plasma on small scales, the steep VSFs suggest that the amount of turbulent kinetic energy on small scales is much less than previous estimations based on a Kolmogorov cascade. Hence the amount of turbulent heating may be much less as well.

In this paper, we analyze the kinematics of 9 cool-core clusters with multiphase filaments observed using the Multi Unit Spectroscopic Explorer (MUSE) and 4 using the Atacama Large Millimeter/submillimeter Array (ALMA). The large sample size now allows us to achieve a deeper and more complete understanding of the motion of the ICM as well as the role of turbulent heating. The paper is structured as follows: we present the data acquisition and our method of calculating VSFs in Section~\ref{sec:dataprocess}. We analyze the VSFs in Section~\ref{sec:res}. In Section~\ref{sec:disc}, we discuss the uncertainties and biases involved in our analysis, different drivers of turbulent motions in cluster centers, and turbulent heating rates in these systems. We conclude our work in Section~\ref{sec:conc}. 
	
\section{Data Processing} \label{sec:dataprocess}

\subsection{Data Acquisition} \label{subsec:dataacq}

We procure our data from the extensive survey of 15 cool-core clusters carried out by \cite{Olivares2019}. Out of the 15 sources, 2 sources (M87 and Abell 2597) have already been analyzed by \cite{Li2020}. In the remaining 13 sources, 9 have been observed with both MUSE and ALMA, while 4 have only ALMA data available. The detailed data description can be found in \citet{Olivares2019}. We only comment on the key aspects of the data extraction process that are relevant to our analyses here.	

MUSE is a second generation instrument for the Very Large Telescope (VLT). It is an optical image slicing IFU with a field-of-view of 1$^\prime\times$1$^\prime$. 
The MUSE data ( ESO programme 094.A-0859(A)) have a spatial sampling of 0.2$^{\prime\prime}$ and a spectral resolution of 1.5~\AA. 
The H$\alpha$ velocity maps used in our VSF analysis are obtained by fitting a single Gaussian profile to all emission lines using the PLATEFIT spectral-fitting routine \citep{Tremonti2004}. An additional Gaussian smoothing kernel was applied to some sources in \citet{Olivares2019}. For consistency, we use the original unsmoothed data for all of our analysis. We have verified that our primary results are not sensitive to the smoothing, likely because all our sources are nearby and bright. For all sources, we only select emission-line flux measurements with signal-to-noise ratio (SNR) $>7$ as in \citet{Olivares2019}. We discuss the effects of noise in more detail in Section~\ref{subsec:uncertbias}. We also apply a velocity error cut ($\sim$5~km~s$^{-1}$, less than 3 times the median error) to eliminate pixels with large uncertainties. As discussed in \cite{Li2020}, this mostly eliminates data towards the edge of the filaments which could either be noise or very faint unresolved gases. Our results are insensitive to the exact value of the velocity error cuts.

The CO molecular gas data comprise of both new and archival ALMA observations of CO(1-0), CO(2-1) and/or CO(3-2) transitions. ALMA maps are made using the ``masked moment" technique described by \citet{Dame2011}. The technique creates a three-dimensional mask that considers spatial and spectral coherence in position-velocity space by smoothing the clean data cube with a Gaussian kernel whose full width at half maximum (FWHM) is equal to the synthesized beam. The velocity maps are then created using this mask on the unsmoothed cube keeping only the regions where the CO lines are detected with a significance greater than 3$\sigma$. When multiple transitions are available, we only use CO(1-0) transition line in our analysis for consistency. The sampling size varies from source to source, but is typically comparable to that used by MUSE for the same target.

\subsection{Data Analysis} \label{subsec:dataanavsf}
	
The theoretical study of steady-state incompressible turbulence is first presented in \citet{Kolmogorov1941a}. Since then, extensive studies have been conducted to extend such theories to compressible (supersonic) turbulence \citep{Boldyrev2002, Kritsuk2007, Federrath2013, Padoan2016}. A VSF is a correlation function that measures the relation between velocities and physical scales. In a Kolmorogov cascade, a two-point VSF of order $p$ defined by $S_p(\ell) \equiv \langle |\delta v(\ell)|^p\rangle$ scales as $S_p(\ell)\propto \ell^{p/3}$, where $\ell$ denotes the distance between the two points, and $\langle|\delta v(\ell)|^p\rangle$ is the mean of the absolute velocity difference to the power of $p$ at this distance. We compute the first-order VSFs for our sources, similar to \cite{Li2020}. For every pair of pixels on the map, we record their spatial separation $\ell$ and velocity difference $\delta v$. We then calculate the average $|\delta v|$ in bins of $\ell$ to obtain the VSF. For a classical Kolmogorov turbulent flow, $S_1(\ell) \equiv \langle |\delta v| \rangle \propto \ell^{1/3}$. For compressible (supersonic) turbulence, the power-law index is $\sim$1/2 \citep{Boldyrev1998}. The velocities used in our analysis are 1D line of sight (LOS) velocities obtained from the line shifts while the positions are projected positions in the 2D plane of the sky. We discuss projection effects in more detail in Section~\ref{subsec:uncertbias}.

\section{Results} \label{sec:res}
	
In this section, we present our first-order VSF analysis of the H$\alpha$ filaments, summarized in Figure~\ref{fig:threepanela}. The leftmost panels in Figure~\ref{fig:threepanela} show the \textit{Chandra} X-ray image of these sources, with the corresponding H$\alpha$ filament flux contours (in red) overlaid to show their spatial correlation. To highlight substructures such as sloshing cold fronts or X-ray cavities in the ICM, we applied unsharp masks to the Chandra images. We smooth the original image using two different scales: the expected cavity size and the size of the cool core. We then subtract one smoothed image from the other, making the surface brightness edges more prominent. The middle panel shows the line-of-sight velocities of the H$\alpha$ filaments. We also divide our maps into inner and outer regions somewhat based on the location of the X-ray bubbles to analyse the correlation between SMBH feedback and turbulence. When possible, we also choose a division such that the numbers of pixels of the inner and the outer regions are similar. The black dashed line in the left and middle panels denotes this boundary between the inner and outer regions. The boundary of the region can sometimes cause artificial features in the VSF that mimic energy injection \citep{Mohapatra2019,Ha2022}. We have experimented with varying the boundary size by $\sim\pm25\%$ in each system. Our main conclusions are not sensitive to the exact choice of this boundary.

\setcounter{figure}{1}
	\begin{subfigure}[h!]
		\centering
		\begin{minipage}[b]{\textwidth}
			\includegraphics[width=\linewidth]{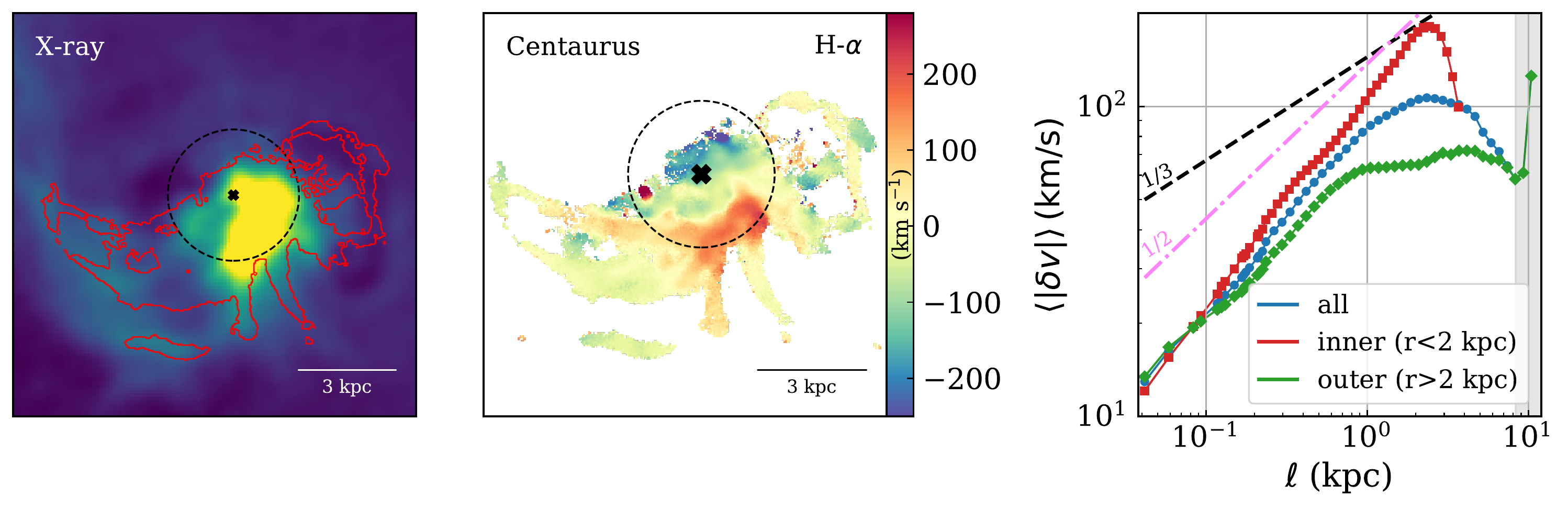}
		\end{minipage}
		\begin{minipage}[b]{\textwidth}
			\includegraphics[width=\linewidth]{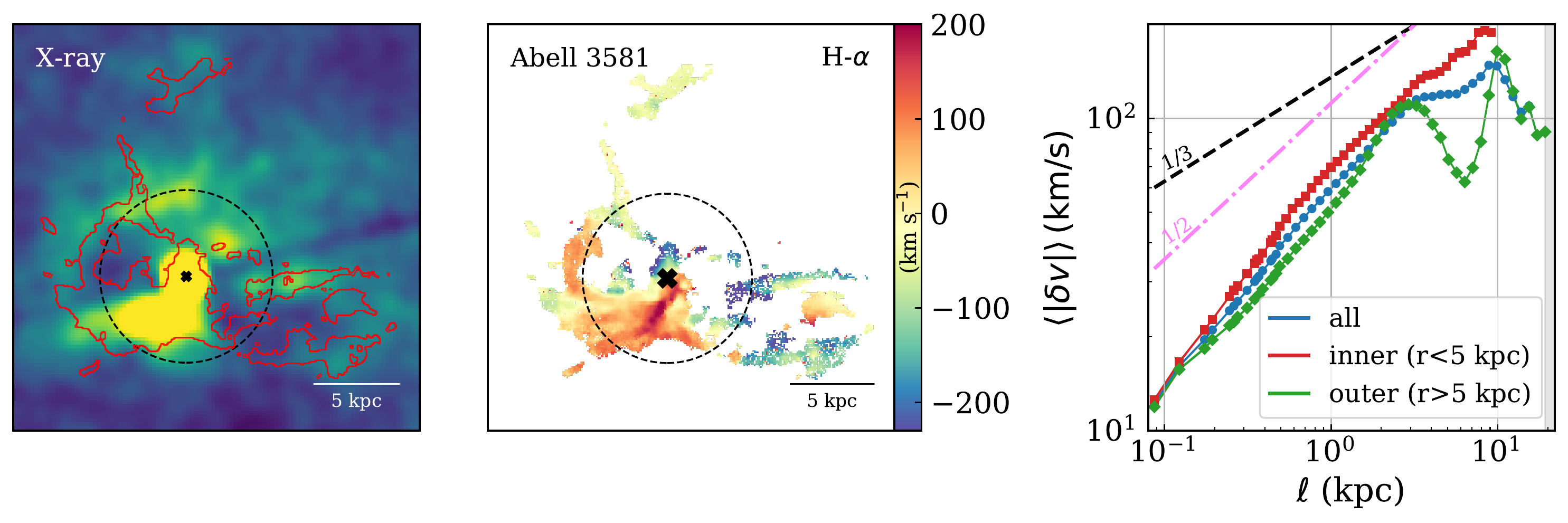}
		\end{minipage}  
		\begin{minipage}[b]{\textwidth}
			\includegraphics[width=\linewidth]{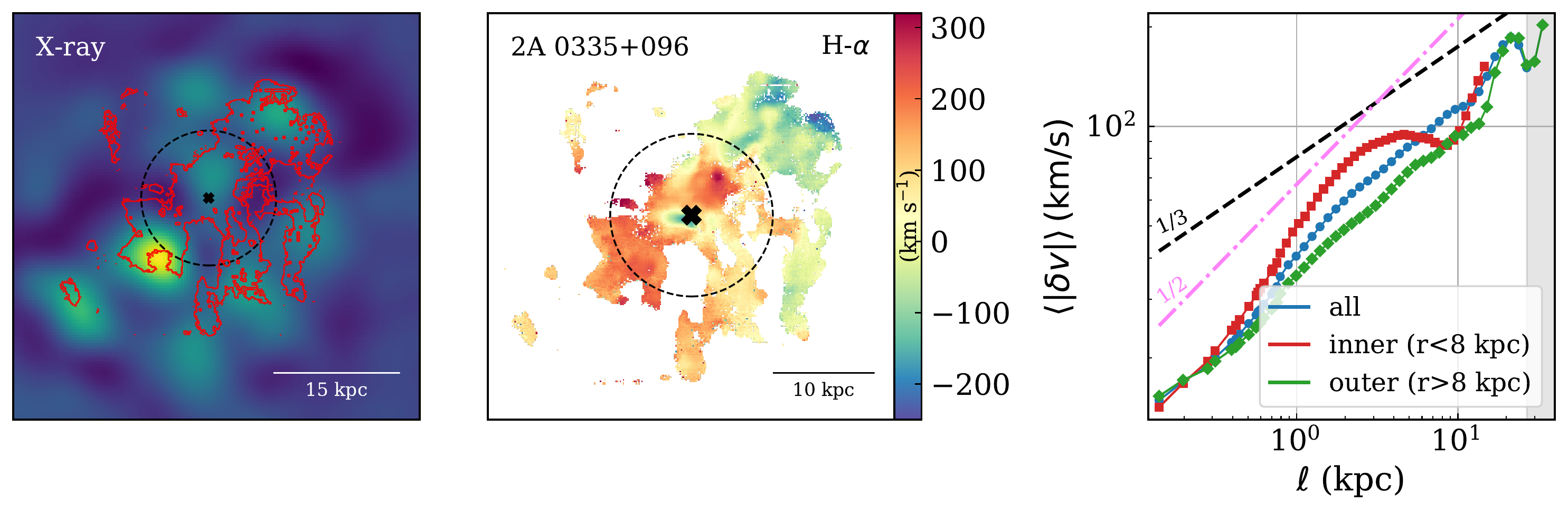}
		\end{minipage}  
		\setcounter{figure}{1}
		\setcounter{subfigure}{-1}
		\caption{\textit{Left panel:} Chandra X-ray unsharp images, with overlaid H$\alpha$ flux contours shown in red. The black `x' marks the center of the BCG where the SMBH is located. The black dashed line marks the boundary for inner and outer filaments. \textit{Middle panel:} The LOS velocity maps of the H$\alpha$ filaments observed by MUSE. The black `x' and dashed line holds the same meaning as in the left panel. \textit{Right panel:} The VSF for all filaments (blue), inner filaments (red) and outer filaments (green), with reference lines of slope 1/3 (Kolmogorov turbulence, shown as black dashed line) and slope 1/2 (supersonic turbulence, shown as pink dot-dashed line). The gray shaded area marks the zone with large uncertainties due to sampling limit, defined as bins with less than 20\% of peak bin size. Each row shows a system in our sample, labelled in the middle panel. Here we show Centaurus, Abell 3581 and 2A 0335+096.}
		\label{fig:threepanela}
	\end{subfigure}

	\renewcommand\thesubfigure{\arabic{subfigure} (Cont.)}
	
	\begin{subfigure}[h!]
		\centering
		\begin{minipage}[b]{\textwidth}
			\includegraphics[width=\linewidth]{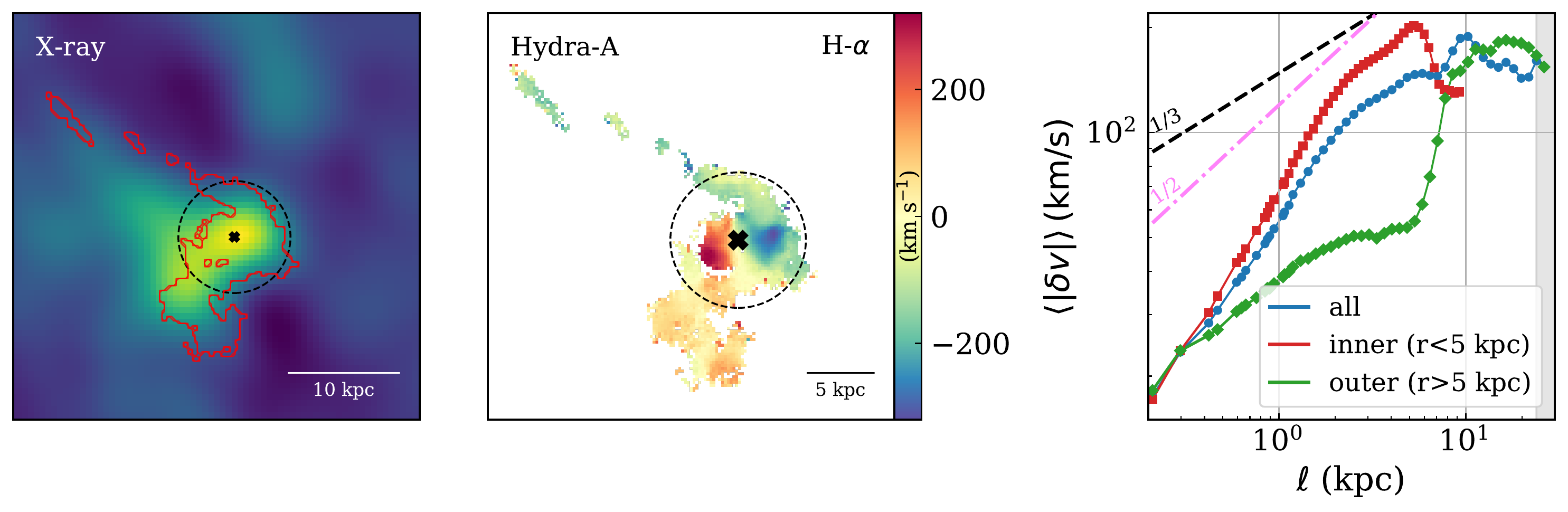}
		\end{minipage}
		\begin{minipage}[b]{\textwidth}
			\includegraphics[width=\linewidth]{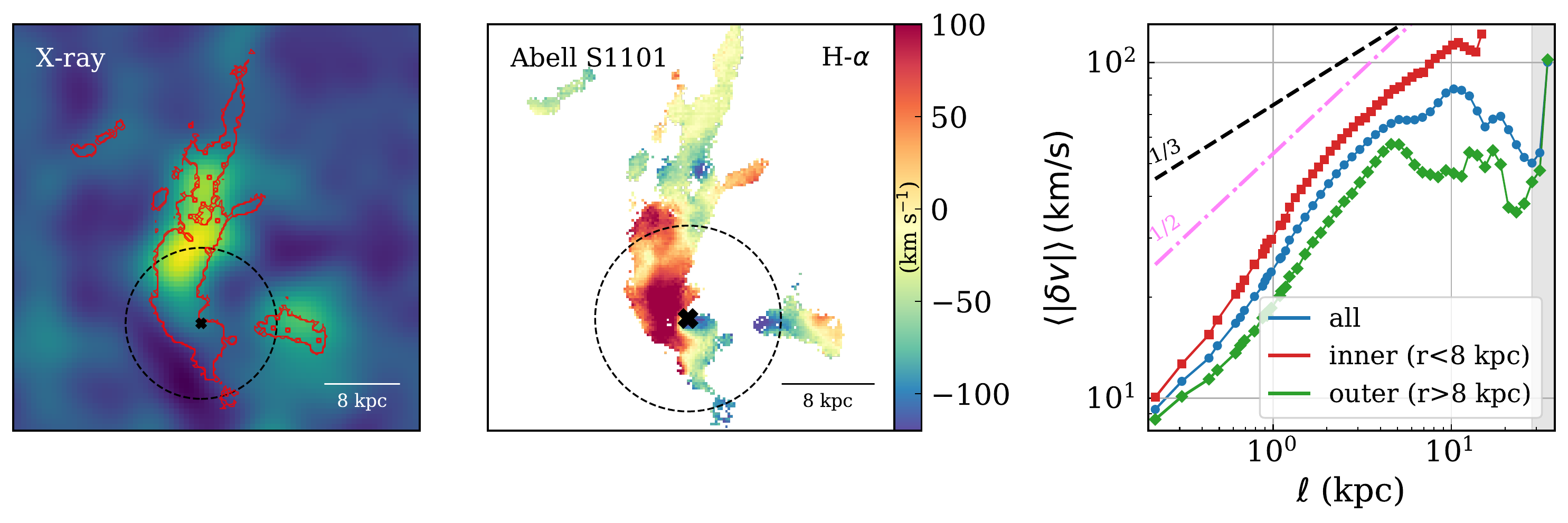}
		\end{minipage}
		\begin{minipage}[b]{\textwidth}
			\includegraphics[width=\linewidth]{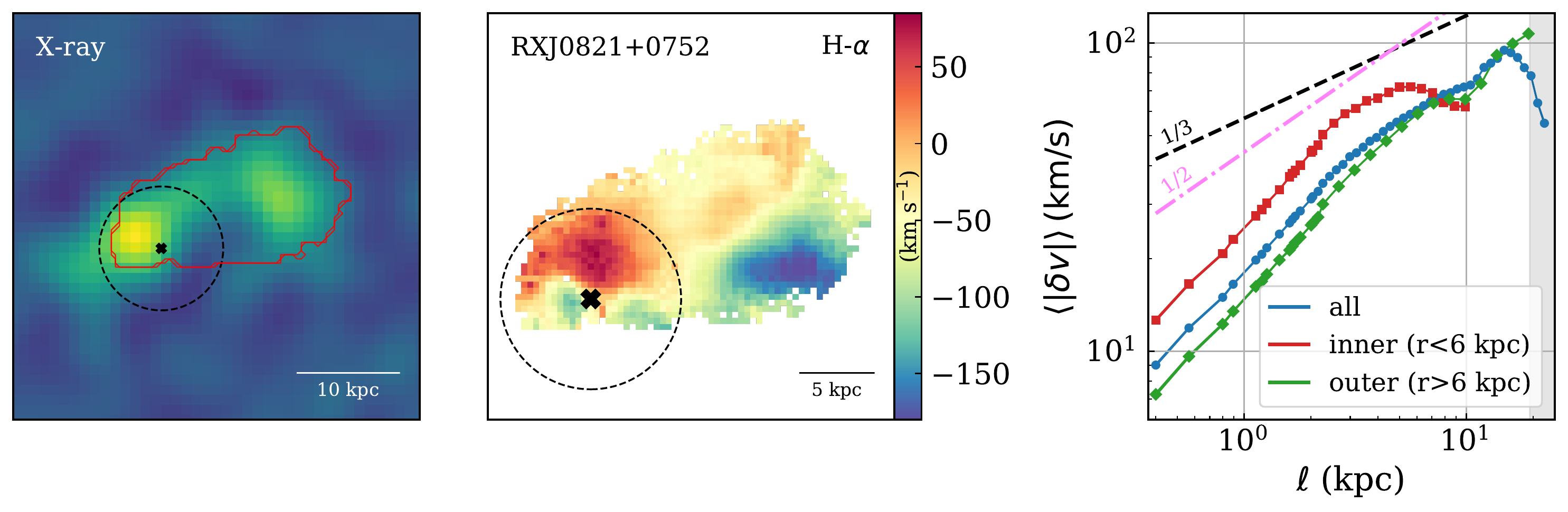}
		\end{minipage} 
		\setcounter{figure}{1}
		\caption{Here we show Hydra-A, Abell S1101 and RXJ0821+0752.}
		\label{fig:threepanelb}
	\end{subfigure}

	\renewcommand\thesubfigure{\arabic{subfigure} (Cont.)}

	\begin{subfigure}[h!]
		\centering
		\begin{minipage}[b]{\textwidth}
			\includegraphics[width=\linewidth]{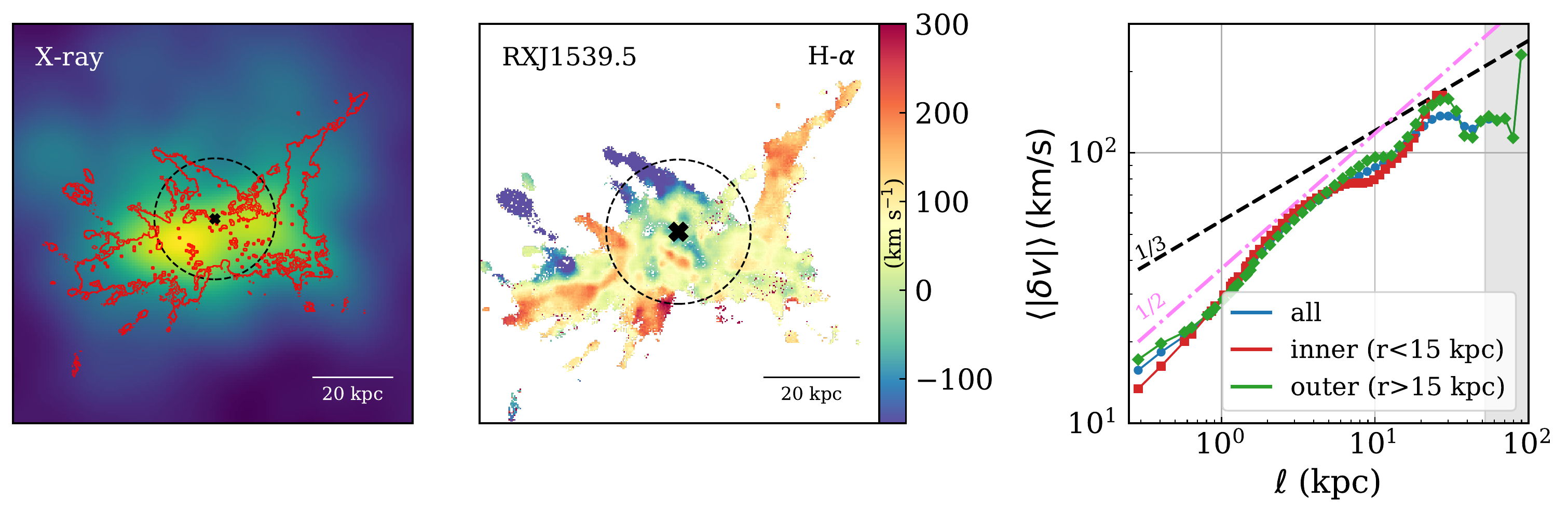}
		\end{minipage}
		\begin{minipage}[b]{\textwidth}
			\includegraphics[width=\linewidth]{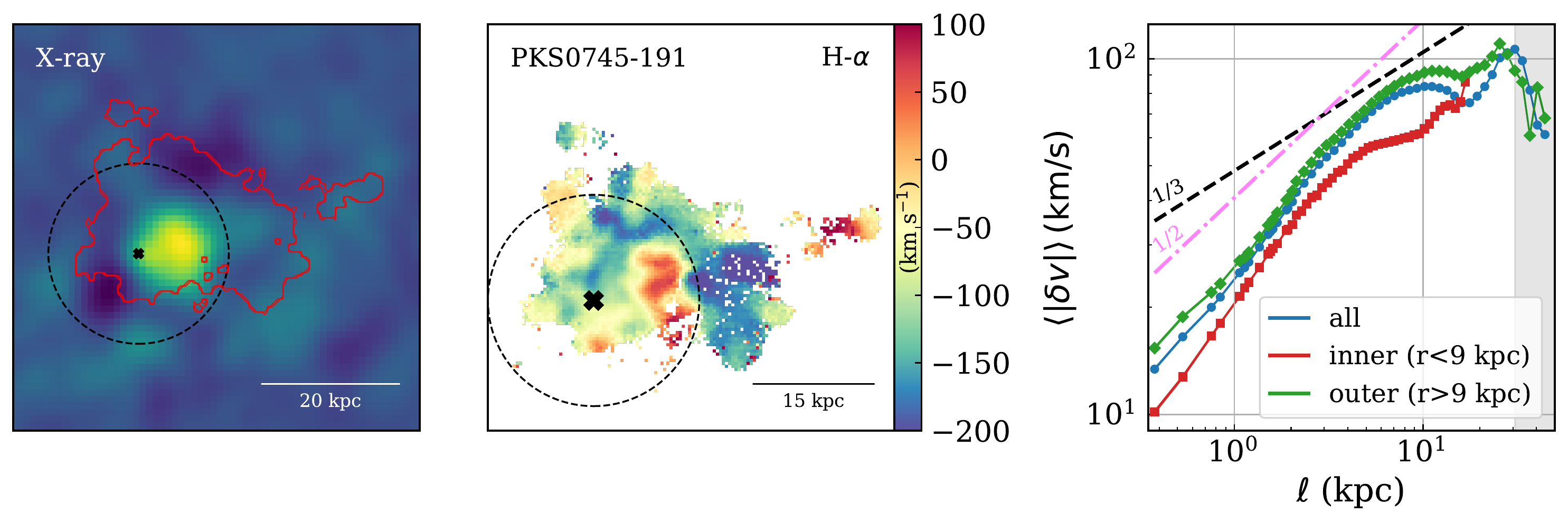}
		\end{minipage}
		\begin{minipage}[b]{\textwidth}
			\includegraphics[width=\linewidth]{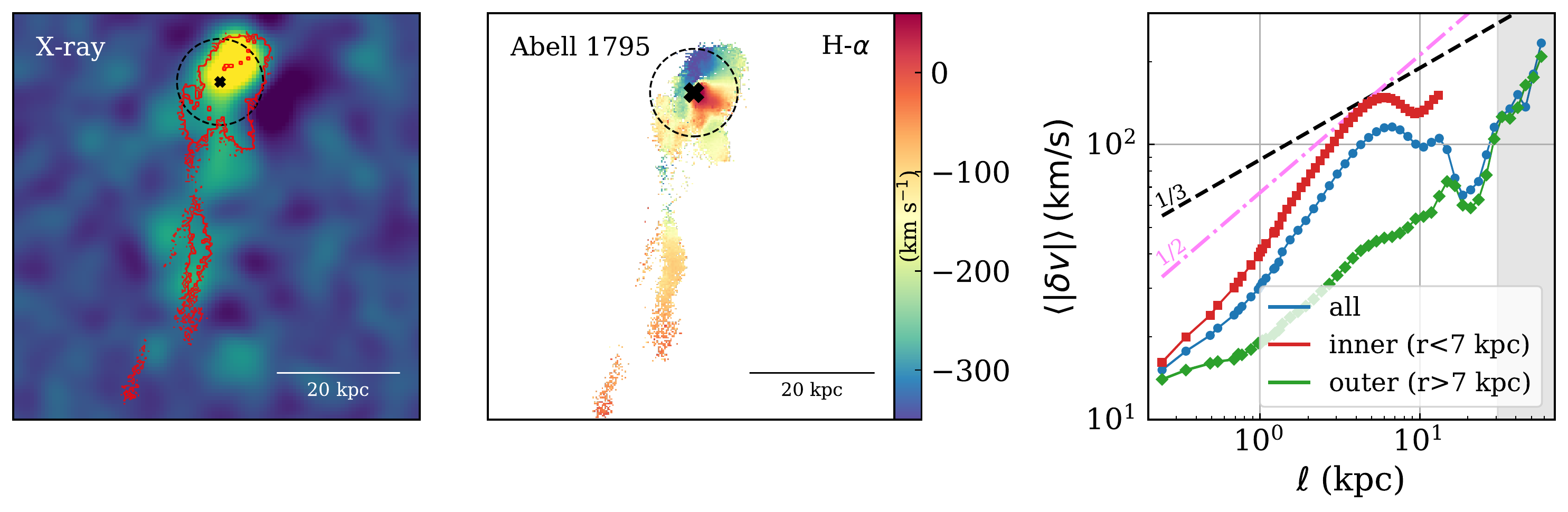}
		\end{minipage}
		\setcounter{subfigure}{0}
		\caption{Here we show RXJ1539.5, PKS0745-191 and Abell 1795.}
		\label{fig:threepanelc}
	\end{subfigure}

The corresponding VSFs are shown in the rightmost panels. 
Most of the VSFs can be described with a single power law on small scales, consistent with the expectation of a turbulent flow. The larger scales show bumps and flattening, suggesting energy injection on these scales. In the following subsections, we categorize the systems based on the unique features revealed in their VSFs. The most common systems have VSFs steeper than Kolmogorov, and the level of turbulence is higher in the inner region, similar to what has been reported in \citet{Li2020}. We refer to these as the vanilla systems and discuss them in Section~\ref{subsec:steepvsf}. Two systems (PKS0745-191 and RXJ1539.5) show the same or even higher level of turbulence in the outer regions, which we discuss in Section~\ref{subsec:out}. In Section~\ref{subsec:flatvsf}, we present the most interesting system Abell 1795, whose VSF shows a robust flattening at small scales, especially in the outer region.

\subsection{The Vanilla Systems} \label{subsec:steepvsf}
 
\noindent
{\bf Centaurus} The VSF of Centaurus shows a steep slope on small scales, and a broad bump from $\sim1-4$ kpc. Chandra X-ray observations of Centaurus have revealed multiple cavities and depressions \citep{2002MNRAS.331..273S, 2005MNRAS.360L..20F, Sanders2016}. Many of them are located outside the spatial coverage of the H$\alpha$ filaments. The energy injection at $\sim 4$ kpc revealed by the VSF may be related to the pair of inner X-ray cavities filled with radio emission \citep{2002MNRAS.334..769T}. \citet{Sanders2016} identify an inner shell-like X-ray structure of 1.9 kpc in radius, which may be a shock driven by SMBH feedback. The structure likely corresponds to the bump in the inner VSF at $\sim 2$ kpc. Galaxy clusters also often show signs of ``sloshing'' caused by sub-structures (smaller clusters or groups) merging with the main cluster. Centaurus is experiencing sloshing motions, but the scales are beyond what can be reliably probed with H$\alpha$ filaments.

\noindent
{\bf Abell 3581} The VSF of Abell 3581 has a steep slope on small scales and reveals multiple driving scales at $\sim 3$ kpc, $\sim 9$ kpc, and possibly larger scales. The first driving scale corresponds to a pair of inner X-ray bubbles filled with radio emission at $\sim 3$ kpc \citep{Johnstone2005}. The second driving scale is likely associated with an outer bubble with a size of 8.4 kpc \citep{Canning2013}. X-ray observations of Abell 3581 also show signs of sloshing motions. The VSF of the outer filaments suggest additional driving at $\sim 20$ kpc or larger, albeit with poor sampling statistics, which may be a result of sloshing. 
    
\noindent
{\bf 2A 0335+096} X-ray observations of 2A 0335+096 reveal multiple cavities (\citet{Sanders2009} report five clear X–ray cavities) as well as multiple X–ray bright blobs. The VSF for 2A 0335+096 has a prominent peak at $\sim20$~kpc, which corresponds to the location of a large X-ray cavity detected in the northwest part of the source \citep{Sanders2009}. There is a hint of additional energy injection on even larger scales. This may be related to older bubbles but may also be a result of merger or sloshing, both of which have been suggested based on X-ray observations \citep{Mazzotta2003, Werner2006}. The VSF of the inner filaments suggest additional energy injection at $\sim 3$ kpc, smaller than any of the cavities identified in \citet{Sanders2009}. It is possible that the jets are directly interacting with the filaments in the center of the cluster, as is suggested by \citet{2010MNRAS.405..898O}. Future X-ray and radio observations may reveal more details of the jet-ICM interaction on small scales.
    
\noindent
{\bf Hydra-A} Hydra-A has strong AGN feedback with multiple generations of X-ray bubbles out to $\sim 200$ kpc  \citep{Wise2007, Nulsen2005}. The VSF shows a prominent bump at $\sim 10$ kpc, which is likely associated with the inner bubbles of sizes $\sim15$~kpc at $\sim30$~kpc. The VSF also suggests energy injection on $\sim 2-5$ kpc. Radio observations reveal several bright knots in the jets at a few kpc scales \citep{1990ApJ...360...41T}. The features in the VSF may be a result of direct interactions between the jets and the ionized gas. The velocity map of the inner filaments shows a clear gradient that likely arises from the rotation of a disk-like structure, which is also observed in CO emissions \citep{2014MNRAS.437..862H}. The rotation can contribute to the steepening of the VSF. The rotation pattern is less prominent in the outer filaments and the corresponding VSF is also flatter. 
	
\noindent
{\bf Abell S1101} Abell S1101 is a system currently experiencing strong AGN feedback based on radio and X-ray observations. The observed radio lobes extend over $\sim 10$ kpc and are spatially correlated with a southern cavity as well as the cool filaments. The cool filaments in the entire system extend over $\sim40$~kpc. The VSFs reveal a driving scale of $\sim 10$ kpc and $\sim 20$ kpc, consistent with the sizes of the X-ray bubbles reported in \citep{Rafferty2006, Werner2011}. The extended flattening from $\sim 8$ kpc down to $\sim 3$ kpc suggest additional energy injection on these smaller scales. Future radio and deeper X-ray observations may confirm the presence of smaller bubbles. Sloshing is a common feature in the X-ray observations of cool-core clusters. The VSF of Abell S1101 shows a convincing decline toward larger scales at $\ell>10$ kpc, suggesting a lack of large-scale energy input. This is consistent with the X-ray analysis which shows a lack of sharp surface brightness discontinuities as sloshing signatures \citep{Werner2011}.

\noindent
{\bf RXJ0821+0752} The VSF of the inner filaments in RXJ0821+0752 reveals a driving scale at $\sim 4-5$ kpc, corresponding to the putative X-ray cavity reported in \citet{Vantyghem2019}. 
The total and outer filament VSFs show additional energy injection on larger scales ($\sim10-20$ kpc). RXJ0821+0752 is experiencing sloshing motion which is supported by the plume-like structure as well as the velocity gradient on large scales. The sloshing motion could be the source of additional energy injection at larger scales ($\sim10-20$~kpc).
 
\subsection{Systems without Higher Central Turbulence} \label{subsec:out}
\noindent
{\bf RXJ1539.5} RXJ1539.5 is the farthest of all the systems in the sample \citep{Olivares2019}. There is no reported detection of X-ray cavities in the literature. The VSF suggests multiple energy injections, at a few kpc, $\sim 20$ kpc, and possibly on even larger scales. Notably, the VSFs of the inner and the outer regions are almost entirely overlapping. This is drastically different from the vanilla systems discussed previously, where the inner VSF has a higher amplitude. 
 
\noindent
{\bf PKS0745-191} The VSF of PKS0745-191 reveals energy injection at $\sim 10$ and $\sim 20$ kpc, corresponding to the cavities observed at these scales \citep{Russell2016}. There may also be additional energy injection from sloshing motion on large scales, suggested by the presence of cold fronts \citep{Sanders2014}. PKS0745-191 is the only system where the amplitude of the VSF of the outer filaments is higher than the inner filaments. The presence of a huge X-ray cavity at larger length scales could explain the higher amplitude of turbulence in the outer filaments. 

On average, one would expect the effect of AGN feedback to be stronger closer to the cluster center where the SMBH is located. This has been shown in both analytical calculations and numerical simulations \citep{2005MNRAS.363..891F, Li2017}. Most of the systems in our analysis and in \citet{Li2020} do follow this expectation. RXJ1539.5 and PKS0745-191 suggest that AGN feedback can be highly variable, and a strong outburst can be followed by a period of low feedback activities. The rarity of systems like RXJ1539.5 and PKS0745-191 indicates that this phase is short-lived. We speculate that the SMBHs in these systems are preparing for the next major outburst and signs of this may be revealed with deeper observations in the future.
	
\subsection{Kolmogorov Slopes on Small Scales} \label{subsec:flatvsf}
		
\noindent
{\bf Abell 1795} Abell 1795 hosts a very long H$\alpha$ filament that extends over $\sim 50$ kpc to the south of the BCG \citep{McDonald2009}. Extensive Chandra X-ray observations reveal multiple cavities as signatures of multiple episodes of AGN activity. For example, \citet{Kokotanekov2018} identify an arc depression at 16 kpc from the center in the north with a radius of 7 kpc, which is the likely causes of the bumps in the VSF at these scales. The VSF of the outer filaments appear to show additional energy injection on larger scales ($>30$ kpc). \citet{Walker2014} identify a large (34 kpc radius) cavity in the north, on the opposite side of the filaments. The filaments are likely showing motion driven by the southern counterpart. It has also been suggested that Abell 1795 is experiencing sloshing \citep{Markevitch2001}. Hence, the additional energy injection on large scales may also partly come from sloshing.

The most remarkable feature in the VSFs of Abell 1795 is that they flatten on small scales. This flattening is especially prominent for the outer filaments. At $\ell>1$ kpc, the VSF of the inner filaments looks very similar to the other systems with a rather steep slope. The VSF for the outer filaments is much shallower, but is still slightly steeper than Kolmogorov. At $\ell<1$ kpc, the slope becomes very close to Kolmogorov. 

The steep slopes of the VSFs found in \citet{Li2020} have been a puzzle. Two of the systems have slopes even steeper than supersonic turbulence. This is seen in many of the systems in our sample as well (see Table~\ref{tab:bpl} for summary). Simulations including magnetic fields do not produce such steep slopes either \citep{Mohapatra2022}. Recent kinetic plasma simulations also show power spectra steeper than Kolmogorov, but the slope is closer to supersonic turbulence within the dynamic range of the simulation \citep{Arzamasskiy2022}. 

	\begin{table}
		\begin{center}
			\begin{tabular}{c |  c |  c | c | c | c | c | c | c}
                System & redshift  & sampling & seeing & $\alpha$ & $\ell_{\rm u}$ & $\ell_{\rm d}$ & $n_e$ & $r_{\rm max}$ \\
    		    &  & (kpc) & (kpc) &  & (kpc) & (kpc) & (cm$^{-3}$) & (kpc)\\
				\hline\hline
				\rule{0pt}{3ex}  Centaurus & 0.01016 & 0.04 & 0.21 & 0.61 & 1 & 2.5 & 0.043 & 6 \\
				\rule{0pt}{2ex}  Abell 3581 & 0.02180  & 0.09 & 0.35 & 0.62 & 2.5 & 3.6 & 0.018 & 13 \\
				\rule{0pt}{2ex}  2A 0335+096 & 0.03634  & 0.14 & 0.77 & 0.53 & 8 & 20 & 0.045 & 20 \\
				\rule{0pt}{2ex}  Hydra-A & 0.05435  & 0.21 & 0.74 & 0.78 & 2 & 5.5 & 0.028 & 21 \\
				\rule{0pt}{2ex}  Abell S1101 & 0.05639  & 0.22 & 0.88 & 0.76 & 2.5 & 6 & 0.022 & 30 \\
				\rule{0pt}{2ex}  RXJ0821+0752 & 0.10900  & 0.4 & 2.4 & 0.78 & 3 & 10 & 0.019 & 20 \\
				\rule{0pt}{2ex}  RXJ1539.5 & 0.07576  & 0.29 & 1.4 & 0.57 & 5 & 25 & 0.012 & 52 \\
				\rule{0pt}{2ex}  PKS0745-191 & 0.10280 & 0.38 & 2.1 & 0.66 & 5 & 10 & 0.051 & 37 \\ 
				\rule{0pt}{2ex}  Abell 1795 & 0.06326 & 0.24 & 1.6 & 0.84 & 4 & 6 & 0.014 & 55 \\ \hline\hline
				\rule{0pt}{2ex}  Abell 262 & 0.01619 & 0.03 & & 0.97  & 1 & 1.7 & 0.016 & 8 \\
				\rule{0pt}{2ex}  Abell 1664 & 0.12797  & 0.69 & & 0.89 & 5 & 8.3 & 0.041 & 10 \\
				\rule{0pt}{2ex}  Abell 1835 & 0.25200  & 1.4 &  & 0.66 & 10 & 22 & 0.12 & 8 \\
				\rule{0pt}{2ex}  Phoenix-A & 0.59600  & 0.68 & & 0.93 & 7 & 12 & 0.2 & 20 \\ \hline\hline
				\rule{0pt}{2ex}  Virgo & 0.00428  & 0.02 & 0.08 & 0.93 & 0.3 & 1.5 & 0.068 & 3 \\
				\rule{0pt}{2ex}  Perseus & 0.01790  & 0.26 & 0.9 & 0.56 & 3 & 10.8 & 0.032 & 30 \\
				\rule{0pt}{2ex}  Abell 2597 & 0.08210  & 0.3 & 1.5 & 0.79 & 4 & 7.7 & 0.036 & 25\\
    \hline
			\end{tabular}
		\end{center}
    \caption{Summary of the systems used in this study. The first two groups are from \citet{Olivares2019}. The first group of systems has been observed with MUSE and is discussed in Section~\ref{sec:res}. The second group has only been observed with ALMA. The last group is from \citet{Li2020}, where Virgo and Abell 2597 are observed by MUSE and Perseus is observed with the optical imaging Fourier transform spectrometer SITELLE at CFHT. Seeing listed here is the average delivered seeing corrected for airmass. $\alpha$ is the slope of the VSF. When fitting the slope, we choose the lower end to be the sampling limit and the upper end slightly below the injection scale ($\ell_{\rm u}$) to allow a single slope. For Abell 1795, we fit a broken power law. The slope between 1 and 4 kpc is listed in the table, while the slope for $\ell<$ 1 kpc is 0.51. The driving scales, $\ell_{\rm d}$, are marked in Figure~\ref{fig:all_vsf}, and are used to compute the driving scale heating rate, discussed in Section~\ref{subsec:qturb}. The electron densities ($n_e$) are obtained from the Archive of Chandra Cluster Entropy Profile Tables (ACCEPT) catalogue \citep{Cavagnolo2009} (https://web.pa.msu.edu/astro/MC2/accept/) except for Phoenix \citep{2019ApJ...885...63M, Kitayama2020}, and they are measured at the outer edge of the cool filamentary structures of each system, $r_{\rm max}$. They are used to compute cooling and heating rates in Section~\ref{subsec:qturb}.}
		\label{tab:bpl}
	\end{table}

The most plausible explanation seems simply that SMBH feedback does not drive volume-filling turbulence efficiently \citep{Zhang2022}. SMBH feedback is highly time-variable, with different generations of bubbles produced over $\sim 10s$ Myr timescales. The eddy-turnover time at the driving scale is of the order of $\sim 100s$ Myr. Thus, steady-state turbulence (to which Kolmogorov's theory applies) cannot be established from the driving scale. The VSFs therefore mainly reflect the motions of the driver (SMBH-driven jets and bubbles) rather than the cascade. 

The main issue with this explanation is that as long as viscosity is suppressed in the ICM \citep[e.g.,][]{Zhuravleva2019}, it is generally assumed that turbulence should develop. There are also other drivers of turbulence in the ICM, such as sloshing and Type Ia Supernovae (SNIa) \citep{LiM2020}. Turbulence driven by these processes is expected to be more volume-filling, so we should still expect a VSF slope close to Kolmogorov on small scales, which is missing in \citet{Li2020}.

Abell 1795 demonstrates that the VSF does have a slope very close to the Kolmogorov expectation on very small scales. However, it requires that we can probe small physical scales in places far from the SMBH. Otherwise, the cascade signature can be completely buried under the driver's behavior. We note that several other systems also exhibit a hint of flattening in their VSF, especially for the outer filaments. These include 2A 0335+096, Centaurus, RXJ 1539.5, and Abell S1101, as well as the outer filaments in Abell 2597 which is noted in \citet{Li2020}. Abell 1795 happens to be a nearby system with one of the most spatially extended H$\alpha$ filaments observed, and therefore, it shows the most convincing flattening in its outer VSF. We show the robustness of the flattening further in Section~\ref{sec:disc}.

\section{Discussion} \label{sec:disc}
	
\subsection{Uncertainties and Biases} \label{subsec:uncertbias}

We discuss the main sources of uncertainties and potential biases in this section, including seeing/smoothing, noise, projection effects, bulk motion, and sampling limit. 

\begin{figure}
	    \centering
	    \includegraphics[width=0.6\linewidth]{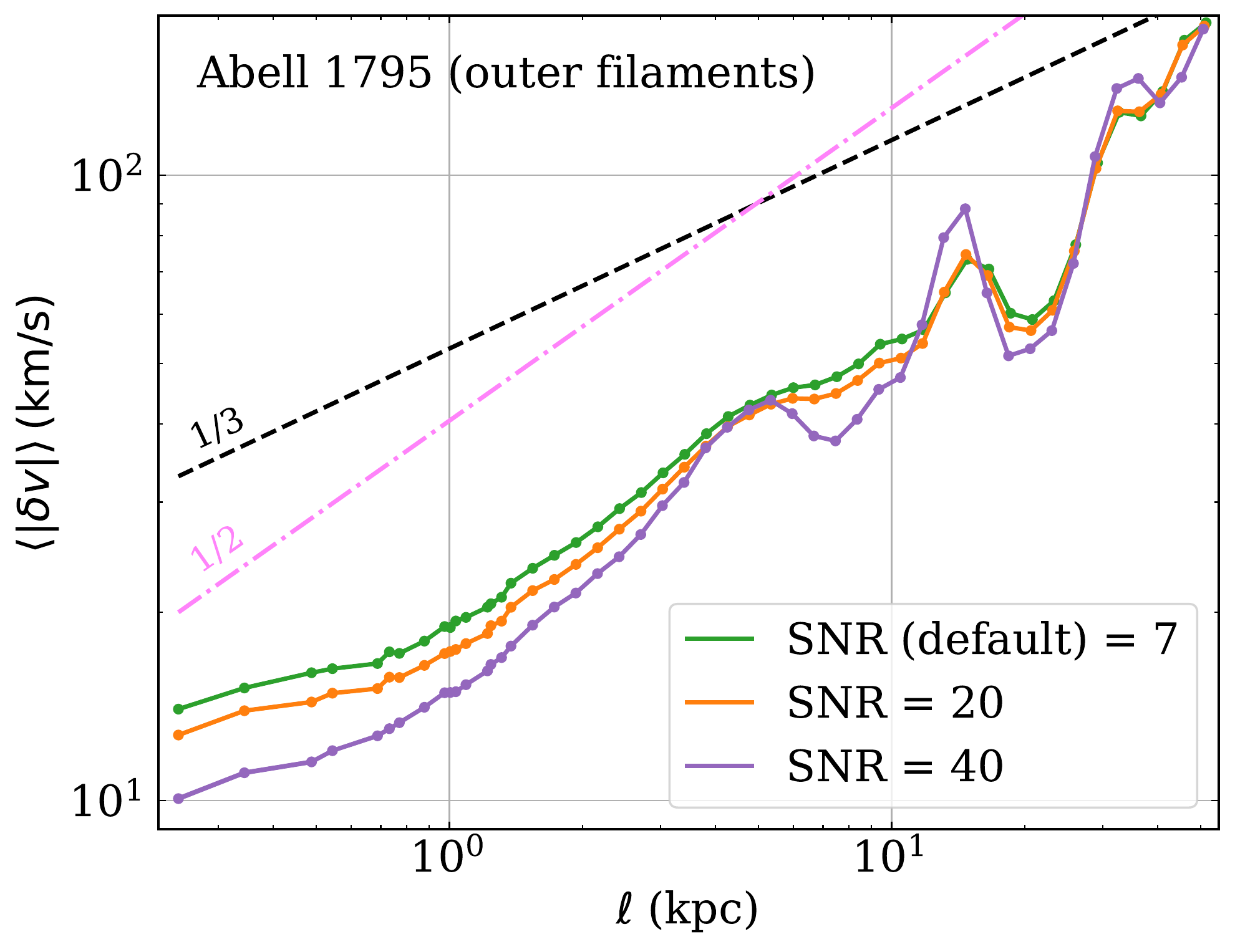}
	    \caption{Comparison of outer filament VSF with different SNR cut applied to the unsmoothed H$\alpha$ map of Abell 1795. The default (blue line) has an SNR cut of 7. The general trend of flattening of VSF at small scales is consistent for all cases, although a stricter SNR cut lowers the VSF amplitude at small scales, likely by removing more noise from the analysis. At larger scales, the VSF is not sensitive to the SNR cuts.}
	    \label{fig:flux_cut}
\end{figure}

Ground-based optical observations can be affected by ``seeing'' due to Earth's atmospheric turbulence. Sometimes, additional smoothing is applied to the observed data to suppress noise, which is usually chosen to have a similar FWHM as the seeing and can have a similar effect. \cite{Chen2022} show that a spatial smoothing function applied to MUSE observations of quasar nebulae can result in a steepening of the VSF at small scales, especially below the FWHM. Since we deal with much brighter sources in this work, the application of smoothing is not essential. We therefore analyze all original data without any additional smoothing applied. This means that our data has more noisy pixels than smoothed data, and noise can flatten VSF on small scales. To understand the effect of smoothing, we have repeated the same analysis for smoothed data. With smoothing, some VSFs show a hint of steepening on small scales, but the overall results are not sensitive to smoothing. The hint of flattening of the VSF in Abell S1101 disappears with smoothing, but for all the other systems discussed in Section~\ref{subsec:flatvsf}, the flattening of the VSF on small scales is robust. Even though we avoid using additional smoothing, it is difficult to remove the smoothing due to atmospheric seeing. Since our VSFs do not show a steepening near the seeing FWHM, the smoothing effect of seeing in our analysis is likely very minor.

To study the effects of noise on our results, we conduct an experiment with the outer filaments of Abell 1795. We compute the VSF of the data selected with different SNR cut. As Figure~\ref{fig:flux_cut} shows, the VSF on large scales remains roughly the same, but a stricter SNR cut results in a smaller amplitude on small scales, as one would expect. This also results in a slight change in the VSF slopes. We obtain slopes of 0.2, 0.2, and 0.27 for $\ell<1$ kpc, and 0.53, 0.57, and 0.67 for $\ell>1$ kpc for experiments with SNR cut at 7, 20, and 40, respectively. 
This experiment also shows that the flattening of the VSF on small scales is robust in Abell 1795.

One of the most fundamental problems that is almost impossible to remove in our analysis is the projection effects. Along each LOS, we may be probing multiple gas clouds with different velocity components. This first projection effect (overlapping clouds along the LOS) can cause a steepening of the VSF \citep[e.g.,][]{Xu2020}, but is likely a rather minor effect in our study. The H$\alpha$ filaments are far from volume-filling, and in the relatively rare cases where a single LOS probes multiple components, the fit is most sensitive to the brightest component. Since there is no reported correlation between the brightness of the filament and its velocity, this does not introduce noticeable bias. 

We can also have clouds that are well-separated in 3D space with high velocity differences, but in the projected plane, they may appear close to each other. The second projection effect can flatten the VSF \citep[e.g.,][]{Qian2015}. As discussed in \citet{Li2020}, this is likely the dominant effect but is difficult to correct for without knowing the three-dimensional distribution of the filaments. We expect the true 3D VSF to be somewhat steeper than the 2D projected VSF we measure here, which does not change our main findings. We summarize the sources of potential biases and their effects in Table~\ref{tab:biases}.

\begin{table}[]
\begin{center}
\begin{tabular}{c | c | c}

sources of biases & effects & severity here  \\
\hline
seeing/smoothing of the data & steepens VSFs below FWHM & minor  \\
\hline
noise  & flattens VSFs & minor \\
\hline
projection (due to multiple LOS components) & steepens VSFs & minor \\
\hline
projection (from the thickness of the structure) & flattens VSFs & may be important \\
\hline
\end{tabular}
\end{center}
\caption{Summary of observational biases and effects on the observed VSF. Note the steepening due to seeing is more severe for shallower VSFs. For steep VSFs like those in this analysis, the effect is a mild suppression of power on all scales with no obvious further steepening on small scales.}\label{tab:biases}
\end{table}

The existence of large-scale bulk motions can also contaminate our analysis. Rotation can add power to the VSF on large scales, which can effectively steepen the VSF \citep{Li2022}. In our sample, most systems do not show large-scale ordered motion except the central region of Hydra-A, which indeed shows a rather steep VSF. In group centrals and isolated elliptical galaxies, many H$\alpha$ structures show a clear rotation pattern \citep{Hamer2016, Olivares2022}. We plan to systematically study the effects of bulk motion in an upcoming work on the kinematics of H$\alpha$ filaments in the lower mass systems. 
 
The sampling limit imposes uncertainties on the VSF at large scales. At large separations, the VSF is averaged over a small fraction of the entire volume. We have shaded (in grey) the range where the number of pairs drops below 20\% of the peak in all our VSF plots to mark the regions with large sampling uncertainties. We avoid over interpreting data at these large scales. 
	
\subsection{Sources of Turbulence in Cluster Centers} \label{subsec:sources}
	
Many physical processes can drive turbulence in cluster centers, including AGN feedback, SNIa and structure formation. Processes related to structure formation, such as mergers, sloshing, and galaxy motions, tend to drive turbulence on scales at or larger than tens of kpc \citep[e.g.,][]{Dolag2005, Vazza2009, ZuHone2013, Shi2018}. Our results show that the kinematics of the filaments in cluster centers are mainly influenced by the activities of the SMBHs, in agreement with the findings in \citet{Li2020}. 
    
The VSFs of several systems show additional energy injection close to the largest scales we can probe (tens of kpc), and the amplitude is typically around 100 km $\rm s^{-1}$. \citet{ZuHone2013} performed MHD simulations and found that sloshing motions in cool-core clusters generate turbulence with $\delta v \sim 50-200$ km $\rm s^{-1}$ on $\sim 50-100$ kpc scales. Thus our measured turbulence on large scales is possibly driven by structure formation processes such as sloshing. However, many AGN-driven bubbles can also reach these scales and velocities. Given the overall spatial extension of the H$\alpha$ filaments, the sampling statistics at tens of kpc scales are rather poor. Future X-ray observations and analyses can potentially help distinguish between sloshing and large bubbles as the main drivers on tens of kpc scales. 

Numerical simulations in \citet{LiM2020} show that SNIa can drive turbulence with velocities of $\sim10-20$ km $\rm s^{-1}$ at $\sim 0.1$ kpc scale in typical massive elliptical galaxy environments. This is smaller than the scales we can probe in our sample except for Centaurus, and may in fact have contributed to the flattening of the Centaurus VSF on small scales. 
Future higher resolution observations with MUSE Narrow Field Mode or ALMA can help measure VSFs on even smaller scales. Numerical simulations tailored for BCGs can also provide more precise theoretical expectations to compare with the observations. 
    
\subsection{Turbulent Heating} \label{subsec:qturb}
	
It is widely accepted that jet mode (radio mode) AGN feedback is crucial in suppressing classical cooling flows in galaxy clusters and maintaining the general quiescent state of massive galaxies in today's universe. However, how AGN feedback operates is far from fully understood. Exactly how the energy of the jets becomes coupled to the surrounding gas is a subject of debate. 

The dissipation of SMBH-driven turbulence can potentially be an important heating mechanism. \citet{Zhuravleva2014} show that in Perseus and Virgo, radiative cooling can be perfectly balanced by the level of turbulent dissipation inferred from surface brightness fluctuation analyses. \citet{Li2020} find that for Perseus, the level of turbulence traced with H$\alpha$ agrees well with those inferred from surface brightness fluctuations and X-ray line-width measurements by Hitomi \citep{Hitomi2016} near the driving scale. However, due to limited spatial resolution, X-ray observations cannot measure the power spectra or the corresponding VSFs of the hot ICM on small scales. If the cool filaments and the surrounding hot plasma remain well-coupled on small scales, then the VSF of the cool filaments reflects the VSF of the hot volume-filling plasma and can be used to infer the level of turbulent heating in these systems. 

\begin{figure}
	\includegraphics[width=\linewidth]{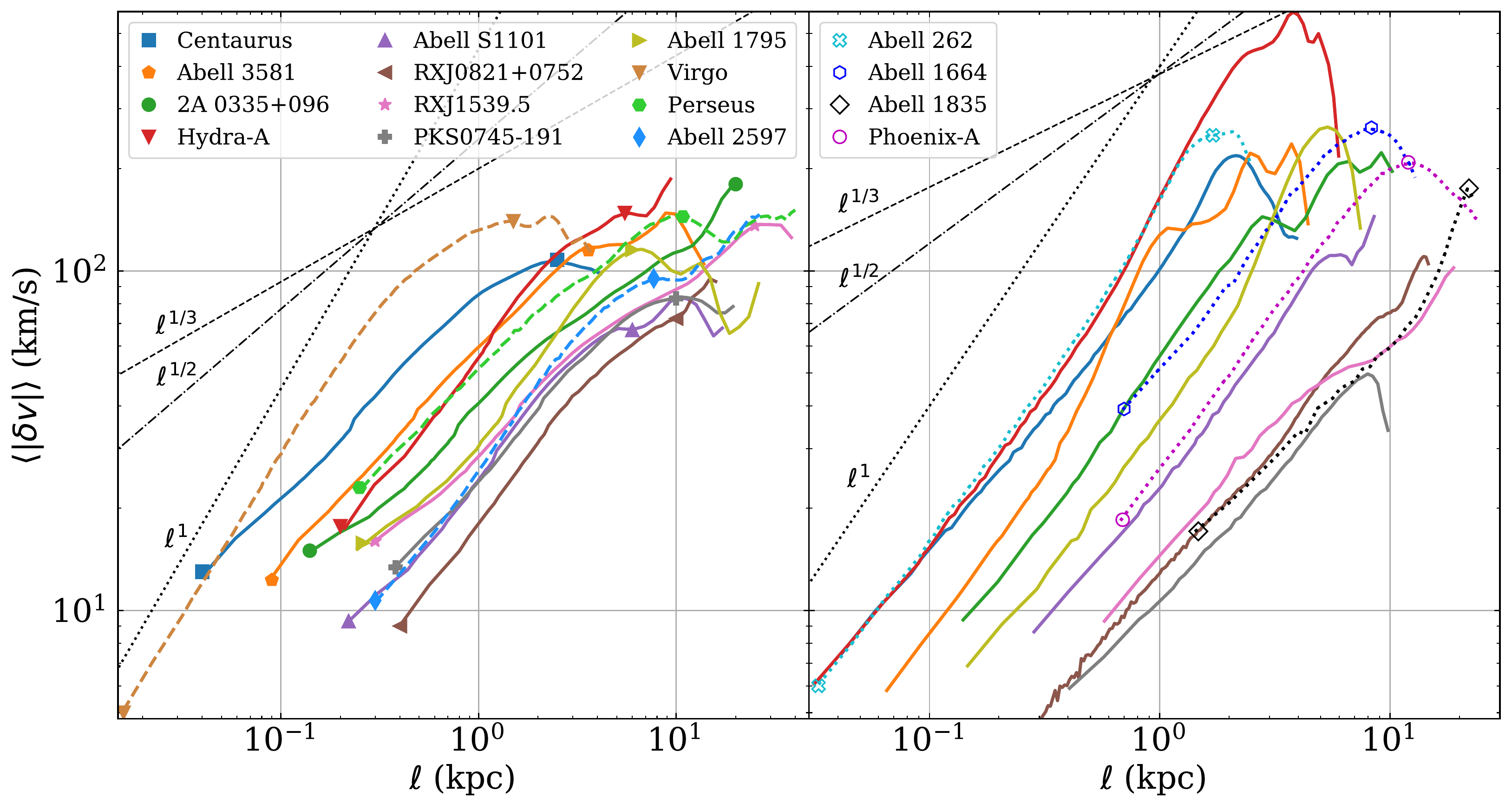}
	\caption{\textit{Left panel:} The VSFs of the H$\alpha$ filaments in the 9 systems analyzed here. Also included are the three systems from \citet{Li2020} in dashed lines. There are two symbols on each curve, marking the sampling limit and the driving scale $l_{\rm d}$ (see Table~\ref{tab:bpl} for summary). These are the two scales where we compute turbulent heating rate (see Section~\ref{subsec:qturb} for details). \textit{Right panel:} The VSFs of the molecular gas observed by ALMA. The 4 labeled systems do not have corresponding MUSE data, and we use the CO VSFs to estimate heating rates at the marked scales.}
	\label{fig:all_vsf}
\end{figure}

Figure~\ref{fig:all_vsf} shows the VSFs of all the sources in our sample. The left panel includes all the systems with MUSE observations that are discussed in Section~\ref{sec:res}. We also include the three systems from \citet{Li2020}, shown in dashed lines. The right panel shows the corresponding VSFs of the molecular gas observed with ALMA. We also include 4 systems from \citet{Olivares2019} that have only ALMA data available. The X-ray and radio maps of these systems are shown in Figure 3 in \citet{Olivares2019}. In most systems, the VSFs of the ionized gas and the molecular component show a reasonably good agreement. The ALMA VSF tends to have a higher amplitude because the detected molecular gas tends to be more centrally concentrated as ALMA is likely filtering out the diffuse, cold molecular gas. As is discussed in Section~\ref{sec:res}, the central filaments have a higher level of turbulence than the outer filaments in most systems. The biases and uncertainties in the ALMA VSF are similar to the H$\alpha$ VSF. The smoothing due to the ALMA beam (of the order of 1'')  is similar to the seeing effect on optical/Halpha images. 
We leave the detailed comparison between the kinematics of the ionized and molecular components for future work. We only use the ALMA data to supplement our analysis for systems that do not have MUSE data available.

For Kolmogorov turbulence, the turbulent heating rate can be estimated as $Q_{\rm turb} \sim \rho v_\ell^3/\ell$, where $v_\ell$ is the velocity at scale $\ell$, and $\rho$ is the gas density. For a steady-state Komogorov flow, $Q_{\rm turb}$ is scale-independent, since $\ell\sim\rho v_\ell^3$.
However, all our sources have VSFs steeper than the Kolmogorov expectation. Therefore, $Q_{\rm turb}$ becomes strongly dependent on $\ell$. 
Since we are only measuring LOS velocities, the 3D $v_\ell$ is related to our measured $v_{\ell,1D}$ as $v_\ell= \sqrt{3}v_{\ell,1D}$, and $Q_{\rm turb} \approx 5.2 \rho v_{\ell,1D}^3/\ell$.
Gas density, $\rho$, is computed as $\rho=\mu m_p (n_i+n_e) \sim \xi \mu m_p n_e$, where $\mu=0.61$ is the mean molecular mass, $m_p$ is the proton mass, and $n_i$, $n_e$ are the ion and electron number densities, related as $n_i=(\xi-1)n_e$ (here $\xi\sim1.912$, for 0.5 solar abundances). 

We compute turbulent heating rate at two scales for each system, using $v_{\ell,1D}$ measured from the VSFs (Figure~\ref{fig:all_vsf}). The first one is computed at the driving scale. When multiple driving scales are present, we choose the smallest. This is usually close to the scale that current X-ray observations can probe. The second one is computed at the smallest scales (sampling limit).
The VSFs of most systems show no signs of flattening even on the smallest scales, suggesting that we are still probing the behavior of the driver rather than the cascade, which will likely only show up on even smaller scales. Given the steep slopes of the VSFs, the true heating rates should be lower than our estimations at the current smallest scales. For these systems, we consider our computed heating rate to be an upper limit. Abell 1795 is the only system where the VSF convincingly transitions to a $\sim$Kolmogorov slope on small scales. This is the only system where we have likely obtained the true heating rate. Several other systems show a hint of flattening on small scales, but it needs to be confirmed by future higher resolution observations. 

\begin{figure}
	\centering
	\includegraphics[width=0.7\linewidth]{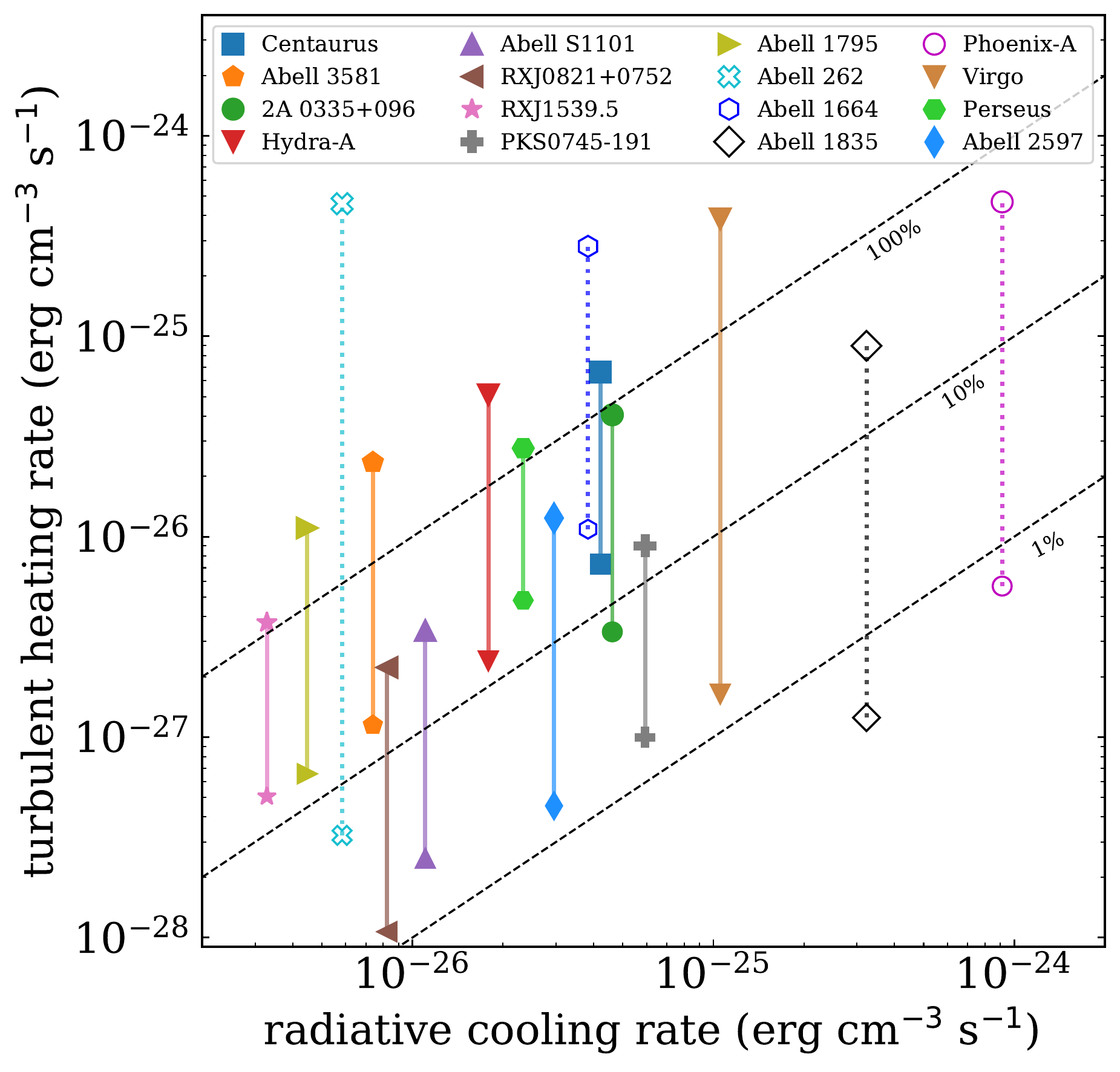}
	\caption{Turbulent heating rate $Q_{\rm turb}$ and the corresponding radiative cooling rate $Q_{\rm cool}$ for all 16 systems where the VSFs of the cool filaments have been measured. 9 of them are systems with MUSE observations discussed in Section~\ref{sec:res}. 4 systems only have ALMA data available (open symbols). The last 3 are from \citet{Li2020}. 
  Each system has two heating rates computed at the driving scale $l_{\rm d}$ and the sampling limit, joined by a vertical dashed line (see $\S\S$\ref{subsec:qturb}). The three reference lines show $Q_{\rm turb}/Q_{\rm cool}=1$, 10\%, and 1\%. Turbulent heating measured at the driving scale can offset a significant fraction of the cooling loss. However, when we resolve turbulence on smaller scales, we find that the actually rate of turbulent heating is much lower due to the steep slopes of the VSFs.}
	\label{fig:qturb}
\end{figure}

We compare the two heating rates and the cooling rate of the cool-cores in Figure~\ref{fig:qturb}. The cooling rates are computed as $Q_{\rm cool} = n_e n_i \Lambda(T)$, where $\Lambda(T)$ is the normalized cooling function. We use $\Lambda(T)$ computed based on \citet{2009A&A...508..751S} assuming 0.5 solar metallicity. Temperatures are measured at the same location as $n_e$.
Cool-core clusters tend to have a steep density gradient in the center. As our analysis shows, the level of turbulence also tends to be higher in the center. Thus in most systems, heating and cooling rates are both declining functions of radius. We use $n_e$ measured at the outer extend of the entire filamentary structure. This is because H$\alpha$ filaments usually only cover the central region of the cool core. $n_e$ measured at the outer edge of the filaments better represents the average core density than if it is measured closer to the center of the cluster. 

Figure~\ref{fig:qturb} shows that for most systems, the heating rate computed at the driving scale is on the same order of magnitude as the cooling rate, consistent with previous analyses using X-ray surface brightness fluctuations \citep{Zhuravleva2014, Zhuravleva2018}. Our computed heating and cooling rates both tend to be higher. This is because \citet{Zhuravleva2018} use volume-weighted gas density for regions typically larger than the spatial extension of the H$\alpha$ filaments. Thus, our $n_e$ is likely higher than theirs. When the heating rate is computed at the smallest scales our MUSE/ALMA observations can probe, the values are typically an order of magnitude lower than those estimated at the driving scales and can only offset order of $10\%$ of the radiative cooling loss. Note that for all systems except Abell 1795, we have likely only obtained an upper limit of the heating rate, as we are still only probing the steep VSF caused by the driver's motion rather than the cascade. 

Our results suggest that turbulent dissipation only contributes to a small fraction of the heating. This is in good agreement with what has been found in numerical simulations \citep[e.g.,][]{Reynolds2015, Yang2016, Li2017, Bambic2018}. Previous analyses based on X-ray observations have inferred a much higher level of turbulent dissipation \citep{Zhuravleva2014, Zhuravleva2018}. This is because current X-ray observations cannot measure the power spectrum on small scales. The previous estimations are effectively the heating rates computed at the driving scales in our study. There are still two possible scenarios allowing turbulent dissipation to be more important than the rates we compute at small scales. One is that some of the steepening in the VSF is due to partial dissipation of turbulence. Another possible scenario is that cool filaments are not well-coupled to the hot plasma on smaller scales. For example, some recent idealized numerical simulations have found that the VSF of the cold phase is slightly lower than that of the hot phase in a multiphase turbulent medium \citep{Gronke2022, Mohapatra2022}. Future X-ray telescopes with higher spatial resolutions can help us better understand the coupling between different phases in the ICM and measure the hot phase turbulence on smaller scales. Note that we refer to heating due to turbulent dissipation as ``turbulent heating'' for simplicity in this work, which does not include heating due to turbulent mixing. 
	
\section{Conclusion} \label{sec:conc}

In this work, we analyze the VSFs of the multiphase filaments in 9 cool-core clusters observed with MUSE and ALMA, as well as 4 observed with ALMA only from \cite{Olivares2019}. For each system, we study the connection between features in the VSFs and potential drivers of ICM turbulent motion, such as SMBH activities, sloshing, and SNIa. We compute turbulent heating rates at large and small scales of the VSFs. Our findings are as follows.

(1) The VSFs suggest that the motions of the filaments are turbulent. There is a good correspondence between the inferred energy injection scales in the VSF and the sizes of SMBH-driven bubbles, suggesting that in the central tens of kpc, ICM turbulence is mainly driven by the activities of SMBHs. Several systems show possible additional energy injection from sloshing on large scales (tens of kpc), but we cannot distinguish between sloshing and older bubbles with our spatial sampling limit on these large scales. 

(2) Most systems show higher level of turbulence in the center and the VSF of the outer filaments shows a lower amplitude, as one would expect. There are two exceptions. In RXJ1539.5, the VSF is almost the same for the entire filament structure. In PKS0745-191, the outer filaments show higher level of turbulence, likely due to a previous powerful AGN outburst suggested by the presence of a huge X-ray cavity. These systems likely represent a special evolutionary stage of the SMBH feedback cycle. 

(3) All our systems have VSFs steeper than the classical Kolmogorov turbulence. Several systems show a flattening of the VSF on the smallest scales, especially for the outer filaments. The most noticeable case is Abell 1795, where the outer VSF convincingly flattens to a Kolmogorov slope. We interpret this as the beginning of the expected Kolmogorov cascade, while the VSFs on larger scales are mainly reflecting the motions of the drivers. This is because SMBH jets and bubbles are very intermittent drivers and cannot establish volume-filling Kolmogorov cascade from the driving scales.

(4) We compare turbulent heating and radiative cooling rates for a combined sample of 16 systems, with 13 from our study and 3 from \citet{Li2020}. The heating rates are computed at both large (the inferred driving scales) and small scales (the sampling limit) of the VSFs. The contribution of turbulent heating can be important when estimated using the driving scale VSFs, which is typically the scale current X-ray telescopes can probe. However, turbulent heating computed on small scales can only offset $\lesssim10\%$ of the radiative cooling loss. We find that heating due to turbulent dissipation is insignificant in cool-core cluster centers, consistent with previous numerical simulations.

Future studies should explore higher-resolution observations using MUSE Narrow Field Mode and/or ALMA, which can potentially probe the VSFs on even smaller spatial scales. Our last conclusion is based on the assumption that multiphase filaments are good kinematic tracers of the hot plasma. This has only been confirmed on large scales observationally. Future X-ray telescopes can help us understand the coupling between different phases on smaller scales. We also require more dedicated numerical studies to better understand the kinematics of the multiphase filaments in galaxy clusters.

\section*{Acknowledgments}

Y.L. acknowledges financial support from NSF grants AST-2107735 and AST-2219686, and NASA grant 80NSSC22K0668. V.O. and Y.S. are supported by NSF grant 2107711, Chandra X-ray Observatory grant GO1-22126X, and NASA grant 80NSSC21K0714. P.G. would like to thank the University Pierre and Marie Curie, the Institut Universitaire de France, the Centre National d'Etudes Spatiales (CNES), the Programme National de Cosmologie and Galaxies (PNCG) and the Physique Chimie du Milieu Interstellaire (PCMI) programs of CNRS/INSU for their financial supports. The study is based on observations collected at the European Organisation for Astronomical Research in the Southern Hemisphere under ESO programme(s) 094.A-0859(A). This work is partly performed at the Aspen Center for Physics, which is supported by National Science Foundation grant PHY-1607611. 

\bibliographystyle{aasjournal}

\expandafter\ifx\csname natexlab\endcsname\relax\def\natexlab#1{#1}\fi
\providecommand{\url}[1]{\href{#1}{#1}}
\providecommand{\dodoi}[1]{doi:~\href{http://doi.org/#1}{\nolinkurl{#1}}}
\providecommand{\doeprint}[1]{\href{http://ascl.net/#1}{\nolinkurl{http://ascl.net/#1}}}
\providecommand{\doarXiv}[1]{\href{https://arxiv.org/abs/#1}{\nolinkurl{https://arxiv.org/abs/#1}}}

\end{document}